\documentclass[prapplied, reprint, superscriptaddress, floatfix, longbibliography]{revtex4-2}

\usepackage{mathtools}
\usepackage{textgreek}
\usepackage{graphicx}
\usepackage{bm}
\usepackage{hyperref}
\usepackage{braket}
\usepackage{xcolor}
\usepackage{multirow}
\usepackage{placeins}
\usepackage{mleftright}
\usepackage[norefs, nocites, ignoreunlbld]{refcheck}
\usepackage{makecell}

\begin{document}

\title{Deterministic Quantum Communication Between Fixed-Frequency Superconducting Qubits via Broadband Resonators}

\author{Takeaki~Miyamura}
\email{miyamura@qipe.t.u-tokyo.ac.jp}
\affiliation{Department of Applied Physics, Graduate School of Engineering, The University of Tokyo, Bunkyo-ku, Tokyo 113-8656, Japan}
\author{Zhiling~Wang}
\affiliation{RIKEN Center for Quantum Computing (RQC), Wako, Saitama 351-0198, Japan}
\author{Kohei~Matsuura}
\affiliation{Department of Applied Physics, Graduate School of Engineering, The University of Tokyo, Bunkyo-ku, Tokyo 113-8656, Japan}
\author{Yoshiki~Sunada}
\affiliation{Department of Applied Physics, Stanford University, Stanford, California 94305, USA}
\author{Keika~Sunada}
\affiliation{Department of Applied Physics, Graduate School of Engineering, The University of Tokyo, Bunkyo-ku, Tokyo 113-8656, Japan}
\author{Kenshi~Yuki}
\affiliation{Department of Applied Physics, Graduate School of Engineering, The University of Tokyo, Bunkyo-ku, Tokyo 113-8656, Japan}
\author{Jesper~Ilves}
\affiliation{Department of Applied Physics, Graduate School of Engineering, The University of Tokyo, Bunkyo-ku, Tokyo 113-8656, Japan}
\author{Yasunobu~Nakamura}
% \email{yasunobu@ap.t.u-tokyo.ac.jp}
\affiliation{Department of Applied Physics, Graduate School of Engineering, The University of Tokyo, Bunkyo-ku, Tokyo 113-8656, Japan}
\affiliation{RIKEN Center for Quantum Computing (RQC), Wako, Saitama 351-0198, Japan}

\date{\today}

\begin{abstract}
Quantum communication between remote chips is essential for realizing large-scale superconducting quantum computers. For such communication, itinerant microwave photons propagating through transmission lines offer a promising approach. However, demonstrations to date have relied on frequency-tunable circuit elements to compensate for fabrication-related parameter variations between sender and receiver devices, introducing control complexity and limiting scalability. In this work, we demonstrate deterministic quantum state transfer and remote entanglement generation between fixed-frequency superconducting qubits on separate chips. To compensate for the sender-receiver mismatch, we employ a frequency-tunable photon-generation technique which enables us to adjust the photon frequency without modifying circuit parameters. To enhance the frequency tunability, we implement broadband transfer resonators composed of two coupled coplanar-waveguide resonators, achieving a bandwidth of more than 100~MHz. This broadband design enables successful quantum communication across a 30-MHz range of photon frequencies between the remote qubits. Quantum process tomography reveals state transfer fidelities of around 79\% and Bell-state fidelities of around 73\% across the full frequency range. Our approach avoids the complexity of the control lines and noise channels, providing a flexible pathway toward scalable quantum networks.
\end{abstract}

\maketitle

\section{INTRODUCTION}
Superconducting quantum computers are one of the most promising platforms for practical quantum computation~\cite{acharya_quantum_2025, lacroix_scaling_2025}. To realize a large-scale superconducting quantum computer, quantum communication between spatially separated quantum processors is essential~\cite{webber_impact_2022, bravyi_future_2022, mohseni_how_2025}. As individual quantum processors approach fundamental limits in cooling capacity~\cite{raicu_cryogenic_2025}, control line density~\cite{krinner_engineering_2019}, and frequency allocation of qubits~\cite{hertzberg_laser-annealing_2021}, modular approaches that connect multiple quantum processing units become increasingly important~\cite{ang_arquin_2024, caleffi_distributed_2024}. This necessity has driven the development of quantum communication protocols for transferring quantum information and generating entanglement between remote superconducting circuits~\cite{cirac_quantum_1997}.

For superconducting quantum computing, two microwave-based approaches have been explored for realizing modular architecture at cryogenic temperatures. One of them exploits discrete resonance modes of a transmission line to mediate interactions between qubits~\cite{zhong_violating_2019, leung_deterministic_2019, chang_remote_2020, burkhart_error-detected_2021, zhong_deterministic_2021, niu_low-loss_2023, qiu_thermal-noise-resilient_2025, heya_randomized_2025, song_realization_2025, mollenhauer_high-efficiency_2025, li_fast_2025}. 
This standing-wave approach has enabled modular quantum computing architectures, but increasing the cable length reduces the free spectral range of the resonance modes, making selective addressing of individual modes difficult.
The other employs propagating microwave photons~\cite{casariego_propagating_2023}. In this approach, quantum information is transferred by converting the sender-qubit excited state into an itinerant microwave photon. This photon then travels along the transmission line to the receiver, where it is absorbed to complete the quantum state transfer. While this approach enables long-distance communication, it has required precise frequency alignment between the interface modes of the sender and the receiver to achieve efficient state transfer.

Various methods for quantum communication using itinerant microwave photons have been demonstrated. Microwave-driven photon-emission schemes with tunable transmons have enabled remote quantum operations~\cite{kurpiers_deterministic_2018, campagne-ibarcq_deterministic_2018, axline_-demand_2018} and have been demonstrated between qubits in separate dilution refrigerators~\cite{magnard_microwave_2020, storz_loophole-free_2023, kulikov_device-independent_2024, storz_complete_2025}. Similarly, transmons equipped with tunable couplers for photon emission have demonstrated quantum communication capabilities~\cite{narla_robust_2016, qiu_deterministic_2025}, including bidirectional multiphoton transfer~\cite{grebel_bidirectional_2024} and chiral photon emission and absorption~\cite{almanakly_deterministic_2025}. However, all previous implementations have relied on frequency-tunable circuit elements, such as flux-tunable transmons, to compensate for fabrication-related parameter variations and ensure resonance matching between sender and receiver. This requirement for tunability introduces additional control complexity through magnetic flux lines, creates potential decoherence channels from flux noise, and poses scaling challenges for large quantum networks~\cite{krantz_quantum_2019}.

In this work, we demonstrate deterministic quantum communication between fixed-frequency superconducting qubits using an itinerant microwave photon. We employ a frequency-tunable photon-generation technique~\cite{miyamura_generation_2025} that adjusts the emitted-photon frequency without modifying circuit parameters, thereby compensating for the fabrication-related frequency mismatch between the sender and the receiver. 
To extend the operational bandwidth of this frequency-tuning approach, we design broadband transfer resonators composed of two coupled coplanar-waveguide resonators, achieving a bandwidth exceeding 100~MHz. 
This design allows us to compensate for the fabrication-related frequency offset of 50~MHz between the sender and the receiver and demonstrate successful quantum communication.
Across the 30-MHz frequency span where the sender and receiver bandwidths overlap, we demonstrate quantum state transfer with process fidelities of approximately 79\% and remote entanglement generation with Bell-state fidelities of approximately 73\%. Our approach eliminates the need for flux-tunable circuit elements and their associated control complexity and noise channels, providing a scalable pathway toward quantum networks based on fixed-frequency qubits.

\section{QUANTUM COMMUNICATION BETWEEN FIXED-FREQUENCY QUBITS}
\begin{figure}
\includegraphics{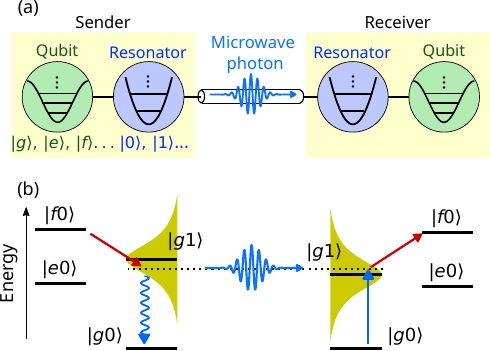}
\caption{Quantum communication between fixed-frequency superconducting qubits with broadband resonators. (a)~Schematic of the system. (b)~Energy-level diagram of the system. A microwave photon is emitted from the sender device (left) via a resonator-assisted Raman transition (red arrow) and absorbed by the receiver device (right). The broadband designs of the communication resonators (yellow Lorentzians) compensate for frequency mismatch, and the frequency-tunable photon generation method~\cite{miyamura_generation_2025} enables quantum communication between fixed-frequency qubits despite such variations.}
\label{figure1}
\end{figure}

\subsection{Basic concept}
Figure~\ref{figure1} shows the quantum communication scheme employed in this work. Two fixed-frequency transmon qubits~\cite{koch_charge-insensitive_2007} are connected via transfer resonators and a transmission line. Here, $\ket{g}$, $\ket{e}$, and $\ket{f}$ denote the three lowest eigenstates of the qubit, while $\ket{0}$ and $\ket{1}$ represent the resonator Fock states. The communication protocol utilizes a resonator-assisted Raman transition~\cite{pechal_microwave-controlled_2014, zeytinoglu_microwave-induced_2015}, enabling the sender device to emit a photon carrying quantum information and the receiver device to absorb it.
Efficient transfer requires precise frequency matching between the transfer resonators, which is challenging due to fabrication-related parameter variations~\cite{li_optimizing_2023,valles-sanclemente_post-fabrication_2023}. We address this frequency-matching requirement by employing a frequency-tunable photon-generation method for fixed-frequency devices~\cite{miyamura_generation_2025}. By off-resonantly driving the $\ket{f0}$--$\ket{g1}$ transition, the emitted-photon frequency can be adjusted to match the transfer-resonator frequency, thereby enabling communication. As long as the frequencies of the transfer resonators lie within their bandwidths, the sender and the receiver can communicate at a matching frequency.

\subsection{Design of the transfer resonator}\label{sec:design}
\begin{figure}
\includegraphics{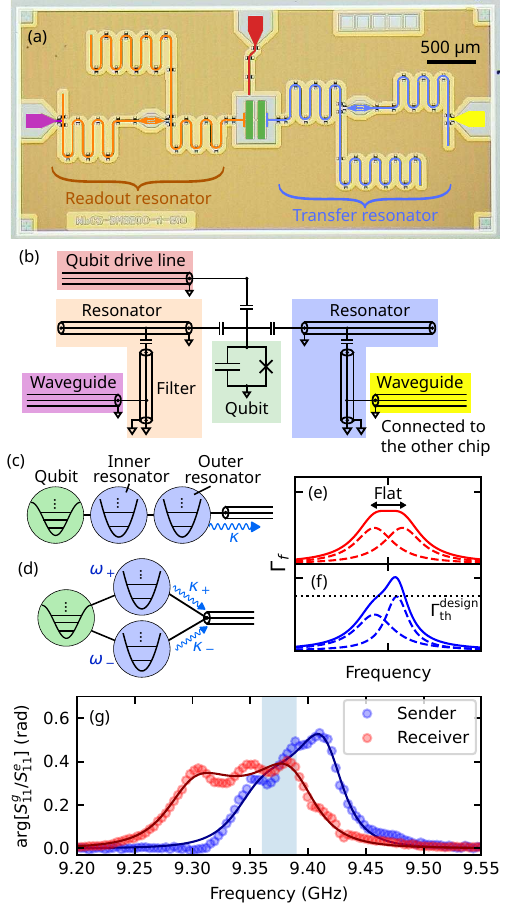}
\caption{(a)~False-colored photograph of the device. A fixed-frequency transmon qubit (green) is capacitively coupled to a readout resonator (orange) and a transfer resonator for single-photon emission and absorption. The qubit is driven via a dedicated control line (red). Both the sender and receiver devices have the same circuit structure. (b)~Equivalent circuit diagram of each device. (c)~Schematic of the qubit--transfer-resonator system. (d)~Equivalent schematic of (c). (e)(f)~Resonator design strategies employed in this work. $\Gamma_f$ is the photon emission rate that corresponds to the resonator spectrum. Dashed curves represent individual resonance modes and solid curves represent the combined resonator spectrum. The black dotted line indicates the threshold of the photon-emission rate used for designing. (g)~Reflection spectroscopy of the transfer resonator. The qubit is prepared in the $\ket{g}$~($\ket{e}$) state, and the reflection coefficient $S_{11,g(e)}$ is measured. Blue~(red) dots represent the phase of the ratio $S_{11,g}/S_{11,e}$ for the sender (receiver) device, with corresponding fits shown as solid lines. The shaded region indicates the frequency range of the emitted photons used in the quantum communication experiments.}
\label{figure2}
\end{figure}

The device used in this work is shown in Fig.~\ref{figure2}(a) and is composed of a fixed-frequency transmon qubit coupled to a readout resonator and a transfer resonator for single-photon emission and absorption.

To enable quantum communication across a wide frequency range, we implement transfer resonators with enhanced bandwidth. Each transfer resonator consists of two coupled coplanar resonators designed to provide a large combined bandwidth.
In the following, we denote the frequency (linewidth) of the first resonance mode of the coupled resonators as $\omega_{+}\,(\kappa_+)$ and that of the second resonance mode as $\omega_{-} \,(\kappa_-)$ [Fig.~\ref{figure2}(d)].
These modes are the eigenmodes obtained through diagonalization of the two coupled resonator modes, excluding the external coupling of the outer resonator.

The two-resonator configuration allows various strategies depending on the design priorities. In this work, we demonstrate two different approaches for the sender and receiver devices. For the receiver device, we choose to prioritize a broad operational bandwidth with a flat spectral response.
To achieve this, we aim to position the two resonator modes with a frequency separation equal to their individual linewidths, i.e., $\omega_+ = \omega_- + \kappa_-$ and $\kappa_+/2\pi=\kappa_-/2\pi=100$~MHz. This configuration creates response of a second-order Butterworth filter that maintains a flat resonator spectrum over an extended frequency range~\cite{thorbeck_high-fidelity_2024}~[Fig.~\ref{figure2}(e)]. 

For the sender device, we choose to maximize the frequency-tunable range while maintaining sufficient photon-emission rate. Following the formalism developed in Ref.~\citenum{miyamura_generation_2025}, the photon-emission rate is defined as the decay rate of the $\ket{f}$ state under the $\ket{f0}$--$\ket{g1}$ drive in the adiabatic limit. When two resonators are contributing to photon emission, the analytical form of this rate $\Gamma_f$ is calculated as
\begin{equation}\label{Gamma_f}
    % \Gamma_f = \frac{4g_{\mathrm{eff}}^2(X^2\kappa_+\delta_{+}+ Y^2\kappa_-\delta_{-})}{4\delta_{+}^2\delta_{-}^2 + (\kappa_+\delta_{+}+\kappa_-\delta_{-})^2},
    \Gamma_f = \frac{g_{\mathrm{eff}}^2J^2\kappa}{\left[(\omega_{\mathrm{ph}}-\omega_{\mathrm{r}})(\omega_{\mathrm{ph}}-\omega_{\mathrm{f}})-J^2\right]^2 + \kappa^2(\omega_{\mathrm{ph}}-\omega_{\mathrm{r}})^2/4}
\end{equation}
% where $\delta_{\pm}$ is the detuning between $\omega_{\pm}$ and the emitted-photon frequency $\omega_{\mathrm{ph}}$, $g_{\mathrm{eff}}$ is the effective $\ket{f0}$--$\ket{g1}$ coupling strength, and $X$~and~$Y$ are the transformation coefficients obtained from diagonalizing the two resonator modes.
where $\omega_{\mathrm{ph}}$ is the emitted-photon frequency, $g_{\mathrm{eff}}$ is the effective $\ket{f0}$--$\ket{g1}$ coupling strength, $\omega_{\mathrm{r}}$~($\omega_{\mathrm{f}}$) is the frequencies of inner (outer) resonator, $J$ is the inter-resonator coupling strength, and $\kappa$ is the external decay rate of the outer resonator [Fig.~\ref{figure2}(c)].
The derivation of Eq.~\eqref{Gamma_f} is provided in Appendix~\ref{app:derivation}. 
% We optimize the resonator parameters to maximize the frequency range over which $\Gamma_f$ exceeds a specified threshold of $\Gamma_{\mathrm{th}}^{\mathrm{design}}/2\pi=8$~MHz [Fig.~\ref{figure2}(f)]. Details on the optimization can be found in Appendix~\ref{sec:design}.
% We note that Eq.~\eqref{Gamma_f} implies a trade-off between the resonator bandwidth and the photon-emission rate. 
Equation~\eqref{Gamma_f} implies a trade-off between the resonator bandwidth and the photon-emission rate.
Physically, a narrower resonator bandwidth increases the density of transmission-line modes accessible from the qubit $\ket{f}$ state near the resonator frequency.
In the adiabatic regime assumed for the photon-shaping technique~\cite{miyamura_generation_2025}, the photon-emission rate is governed by Fermi's golden rule and is therefore enhanced by this increased density of states.
Consequently, a reduction in the resonator bandwidth improves $\Gamma_f$, at the cost of a narrower operational bandwidth.
In Appendix~\ref{sec:design}, we analyze this trade-off and show that the fabricated sender device operates near the optimal point in the parameter space [Fig.~\ref{figure2}(f)].

% Both devices are targeted to the same resonance frequencies and are designed to maximize spectral overlap so as to maximize their spectral overlap.
The transfer-resonator spectrum of each device is obtained with the setup described in Fig.~\ref{figure_wiring} by performing reflection measurements with the qubit in that device prepared in $\ket{g}$ and $\ket{e}$, yielding reflection coefficients $S_{11}^g$ and $S_{11}^e$, respectively.
The resonator spectrum is extracted from the ratio $S_{11}^e/S_{11}^g$, which isolates the spectrum of the target resonator while canceling the response of the other device.
Figure~\ref{figure2}(g) presents the measured transfer-resonator spectra for both devices. 
Both resonators exhibit a bandwidth exceeding 100~MHz.
The bandwidth of the sender device is made narrower in order to maintain the photon-emission rate above a threshold. 
The receiver device achieves a broader linewidth of approximately 150~MHz and a flatter resonator spectrum as intended by the design strategy. 
Despite the fabrication-related frequency offset of approximately 50~MHz between the sender and the receiver, the spectral overlap between the two devices indicates the capability of performing quantum communication. 

If only one resonator were used to realize a photon-emission bandwidth of 100~MHz, the corresponding linewidth of the resonator would have enhanced the energy relaxation of the qubit through off-resonant Purcell effect. By using the two resonator configuration, we provide a band-pass filtering effect that suppresses unwanted decay at the qubit frequency~\cite{sete_quantum_2015, jeffrey_fast_2014}. To further mitigate the Purcell decay, we implement intrinsic Purcell filters~\cite{sunada_fast_2022, sunada_photon-noise-tolerant_2024, spring_fast_2025}, which strongly suppress the coupling between the qubit and the transmission line at the qubit frequency while maintaining efficient photon emission and absorption at the transfer-resonator frequencies. As a result, the qubit energy relaxation times average around 20~$\mathrm{\mu s}$ (Table~\ref{tab_parameters}), which is sufficient for demonstrating quantum communication protocols.
\begin{figure*}[t]
    \centering
    \includegraphics{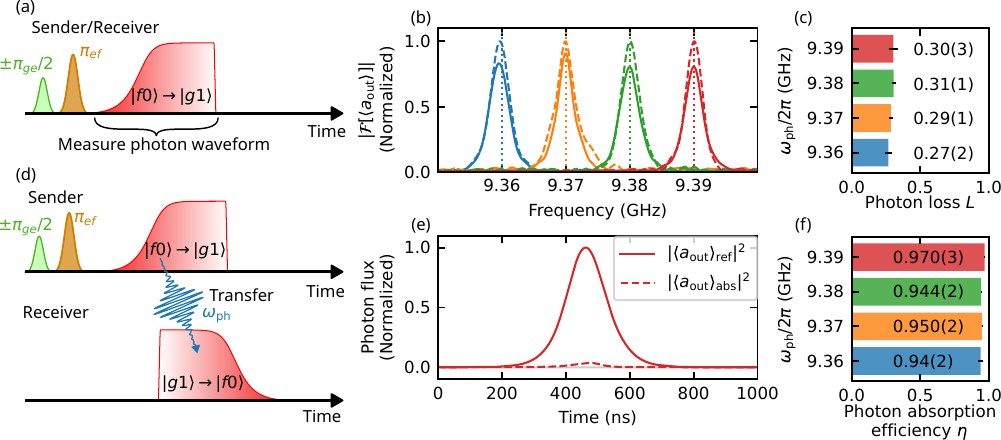}
    \caption{Propagation loss and absorption efficiency. (a)~Pulse sequence to estimate photon loss during propagation. A microwave photon is emitted from the sender and receiver devices separately. The presence of a circulator in the connecting cable (see Fig.~\ref{figure_wiring} in Appendix~\ref{app:setup}) allows an independent measurement of photon loss during propagation and absorption efficiency at the receiver~\cite{kurpiers_deterministic_2018}. (b)~Fourier amplitudes of the emitted photons at each target frequency $\omega_{\mathrm{ph}}$. Solid curves represent photons emitted from the sender device, while dashed curves represent photons emitted from the receiver device. Vertical dotted lines indicate the target photon frequencies $\omega_{\mathrm{ph}}$. (c)~Estimated photon loss for each photon frequency $\omega_{\mathrm{ph}}$. (d)~Pulse sequence for measuring the photon-absorption efficiency. A microwave photon emitted from the sender is absorbed at the receiver using a time-reversed drive pulse. (e)~Measured photon flux at $\omega_{\mathrm{ph}}/2\pi = 9.39$~GHz. The solid curve shows the photon flux without absorption, and the dashed curve shows the residual photon flux after the absorption at the receiver. (f)~Measured absorption efficiency for each photon frequency $\omega_{\mathrm{ph}}$.}
    \label{fig:3}
\end{figure*}
\section{EXPERIMENT}
\begin{figure*}[t]
    \centering
    \includegraphics{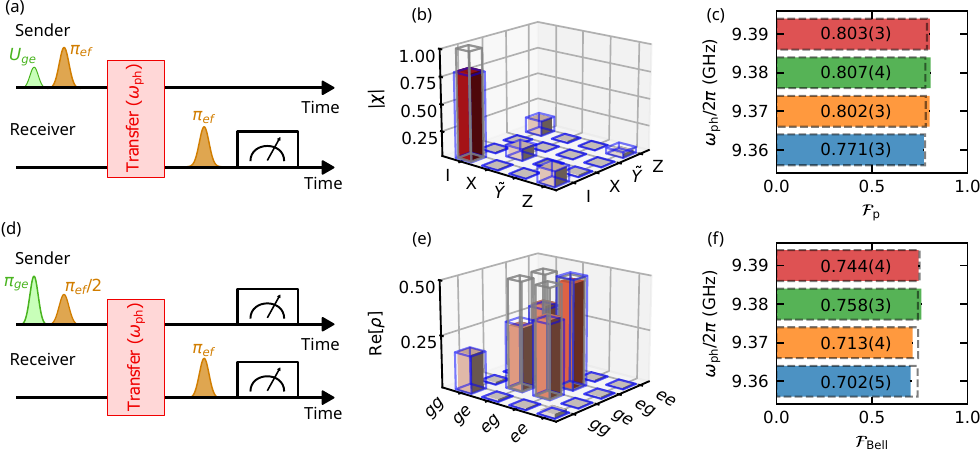}
    \caption{Quantum state transfer and remote entanglement generation. (a)~Pulse sequence for quantum state transfer between the sender and receiver qubits. (b)~Process matrix for quantum state transfer at $\omega_{\mathrm{ph}}/2\pi = 9.39$~GHz. Here, $I,\,X,\,Y$ and $Z$ are the Pauli operators, and $\tilde{Y}=-iY$. Gray and blue wireframes indicate the ideal process and the simulated result, respectively. (c)~Process fidelities for quantum state transfer at each photon frequency $\omega_{\mathrm{ph}}$. The dashed frames represent the simulated results. (d)~Pulse sequence for remote Bell-state generation. (e)~Reconstructed density matrix of the two-qubit Bell state at $\omega_{\mathrm{ph}}/2\pi = 9.39$~GHz. Gray and blue wireframes indicate the ideal state and the simulated result, respectively. (f)~Bell-state fidelities at each photon frequency $\omega_{\mathrm{ph}}$. The dashed frames indicate the simulated results.}
    \label{fig:4}
\end{figure*}
To verify the broadband operation of our system, we demonstrate quantum communication at photon frequencies of $\omega_{\rm ph}/2\pi = 9.36, 9.37, 9.38$, and $9.39$~GHz, all within the spectral overlap region of the transfer resonators shown in Fig.~\ref{figure2}(g). The target photon waveform is set to $\psi_{\rm ph}^{\rm target}(t) = \sqrt{\kappa_{\mathrm{ph}}/2}\,\mathrm{sech}(\kappa_{\rm ph}t)$ with $\kappa_{\rm ph}/2\pi = 2$~MHz.
The hyperbolic secant waveform is chosen because its exponentially decaying tails yield a finite photon-emission rate $\Gamma_f(t)$.
Prior to each measurement, an active reset protocol~\cite{magnard_fast_2018} is applied to initialize both qubits, reducing the residual excited-state population to around 0.3\%.

\subsection{Photon emission and absorption at multiple photon frequencies}\label{sec:absorption}
First, we estimate the photon loss during propagation through the transmission line.
The loss can be determined by emitting a single photon individually from either the sender or receiver device, exploiting the presence of a circulator in the connecting cable~\cite{kurpiers_deterministic_2018} (see Fig.~\ref{figure_wiring} for the measurement setup).
The pulse sequence for this measurement is shown in Fig.~\ref{fig:3}(a).
To observe the photon waveform $\langle a_{\mathrm{out}}(t)\rangle$, we prepare the qubit in the superposition state $({\ket{g}\pm\ket{f}})/{\sqrt{2}}$.
The emitted microwave field is amplified using a flux-driven Josephson parametric amplifier~(JPA)~\cite{yamamoto_flux-driven_2008}.
We perform this measurement $2 \times 10^5$ times for each state and average the results to obtain the photon waveform~\cite{miyamura_generation_2025}.

This approach assumes that both devices emit photons with equal efficiency. However, characterization of our devices reveals deviations from this assumption. 
In particular, the receiver device sometimes exhibits an unexpectedly large decay from the $\ket{f}$--$\ket{e}$ transition and leakage to states beyond $\ket{f}$ during photon emission (see Appendix~\ref{app:qpop}).
To account for this qubit-population imbalance, we measure the qubit population after the photon emission.
Assuming that the population in $\ket{g}$ ($P_g$) after the emission equals the average emitted photon number, we correct for the population imbalance by normalizing the observed photon waveform by $\sqrt{P_g}$.

Figure~\ref{fig:3}(b) shows the Fourier transforms of the emitted photons.
The measured frequencies of the emitted photons from both devices align well with the target photon frequencies.
The photon loss is quantified using the relation
\begin{equation}\label{eq:photon_loss}
    L = 1-\frac{\int |\langle{a_{\mathrm{out}}}(t)\rangle_{\mathrm{tx}}|^2\,dt}{\int |\langle{a_{\mathrm{out}}}(t)\rangle_{\mathrm{rx}}|^2\,dt},
\end{equation}
where $\langle a_{\mathrm{out}}(t)\rangle_{\mathrm{tx(rx)}}$ is the expectation value of the field operator on the transmission line when emitting a photon from the sender (receiver).
We note that this method assumes comparable $\ket{g}$--$\ket{f}$ coherence for both devices. Numerical simulations using the measured coherence times (Table~\ref{tab_parameters}) indicate that the systematic error in $L$ arising from the difference in dephasing rates between the two devices is approximately 1\%, which is within the measurement uncertainty. 
The measured photon loss at each frequency is presented in Fig.~\ref{fig:3}(c).
The loss shows insignificant frequency dependence across the measured range, yielding an average value of 29\%, which is used in the numerical simulations in Sec.~\ref{sec:comm}.

To characterize the absorption efficiency, we perform photon emission and absorption measurements at the same set of frequencies.
The pulse sequence for photon absorption is shown in Fig.~\ref{fig:3}(d).
Since the absorption process is the time-reversed process of emission, the drive at the receiver is the time-reversed drive used for photon emission in Fig.~\ref{fig:3}(a)~\cite{cirac_quantum_1997}.
The delay between the two $\ket{f0}$--$\ket{g1}$ drives is optimized by sweeping the delay time and selecting the value that maximizes the population in $\ket{f}$ state of the receiver qubit~\cite{magnard_microwave_2020}.
Following photon emission from the sender, we either enable or disable the absorption at the receiver and monitor the reflected waveform through the circulator.
When the absorption is enabled, the reflected waveform corresponds to the photon flux that is not absorbed by the receiver, whereas without the absorption, it corresponds to the total photon flux arriving at the receiver.
Figure~\ref{fig:3}(e) presents the measured photon flux at $\omega_{\mathrm{ph}}/2\pi = 9.39$~GHz with and without the absorption.
The absorption efficiency $\eta$ is calculated as
\begin{equation}\label{eq:absorption_eff}
    \eta = 1 - \frac{\int |\langle{a_{\mathrm{out}}}(t)\rangle_{\mathrm{abs}}|^2\,dt}{\int |\langle{a_{\mathrm{out}}}(t)\rangle_{\text{ref}}|^2\,dt},
\end{equation}
where $\langle{a_{\mathrm{out}}}(t)\rangle_{\mathrm{abs(ref)}}$ represents the expectation value of the field operator with (without) the absorption at the receiver.
The absorption efficiency at each photon frequency is shown in Fig.~\ref{fig:3}(f) and is approximately 95\% across the measured frequency range.

\subsection{Deterministic quantum state transfer and generation of remote entanglement}\label{sec:comm}

Next we demonstrate quantum state transfer and remote entanglement generation between the fixed-frequency qubits.

The pulse sequence for quantum state transfer is shown in Fig.~\ref{fig:4}(a).
We prepare six cardinal qubit states at the sender: $\ket{\psi}= \ket{g},\, \ket{e},\, \ket{\pm},\, \ket{\pm i}$, where $\ket{\pm} = (\ket{g} \pm \ket{e})/\sqrt{2}$ and $\ket{\pm i} = (\ket{g} \pm i\ket{e})/\sqrt{2}$.
Following the state preparation, we transfer the quantum state from the sender to the receiver via a microwave photon at frequency $\omega_{\mathrm{ph}}$, followed by an application of a $\pi_{ef}$ pulse at the receiver to complete the protocol.
Quantum process tomography~\cite{chuang_prescription_1997} is performed by preparing the input states at the sender and measuring the output states at the receiver.
Figure~\ref{fig:4}(b) presents the reconstructed process matrix $\chi$ when the transfer is mediated by a photon at $\omega_{\mathrm{ph}}/2\pi = 9.39$~GHz.
The process fidelity $\mathcal{F}_{\mathrm{p}}$ for state transfer using photons at various frequencies is shown in Fig.~\ref{fig:4}(c), achieving values around 79\%.
The consistent fidelities across the 30-MHz frequency range demonstrate that the communication protocol performs similarly regardless of the specific photon frequency within the operational bandwidth.

To understand the sources of infidelity, we perform numerical simulations (see Appendix~\ref{app:sim}) using the measured photon loss, absorption efficiency from Sec.~\ref{sec:absorption}, and the device coherence times listed in Table~\ref{tab_parameters}.
The simulated results, shown as blue frames in Figs.~\ref{fig:4}(b)~and~(e), agree well with the experimental data.
The analysis of the simulation reveals that the infidelity originates from photon loss during propagation ($\approx$~14\%), imperfect photon absorption at the receiver ($1.4\text{--}2.7$\%), qubit energy relaxation ($\approx$~1.7\%), and qubit dephasing~($\approx$~2.6\%).

In addition to quantum state transfer, we demonstrate remote entanglement generation.
The pulse sequence for creating a Bell state between the remote qubits is shown in Fig.~\ref{fig:4}(d).
We perform two-qubit state tomography to characterize the generated entangled state.
Figure~\ref{fig:4}(e) displays the reconstructed density matrix of the two-qubit system when the communication is mediated by a photon at $\omega_{\mathrm{ph}}/2\pi = 9.39$~GHz.
The Bell-state fidelity $\mathcal{F}_{\mathrm{Bell}}$ for entanglement generation using photons at different frequencies is summarized in Fig.~\ref{fig:4}(f).
The fidelities are approximately constant across the measured frequency range, with an average of 73\%.
However, slightly lower fidelities are observed for communication mediated by photons at $\omega_{\mathrm{ph}}/2\pi = 9.36$ and 9.37~GHz.
We attribute this reduction to temporal fluctuations in qubit coherence during these measurements.
Repeated characterization of our devices reveals occasional degradation of coherence times, with values as low as $T_{1,ge} \sim 10\mathrm{\, \mu s}$ and $T_{1,ef} \sim 3\mathrm{\, \mu s}$~(see Appendix~\ref{app:setup}), compared to the typical values listed in Table~\ref{tab_parameters}.
The origin of these coherence time fluctuations is not definitively established.
We suspect coupling to a two-level-system~(TLS) defect, as we occasionally observe sudden shifts in qubit frequency with these devices, a characteristic signature of TLS effects~\cite{muller_interacting_2015, klimov_fluctuations_2018, thorbeck_two-level-system_2023}.
Indeed, numerical simulations including the degraded coherence times yield infidelity contributions from photon loss~($\approx$~15\%), imperfect photon absorption~($1.5\text{--}3.0\%$), qubit energy relaxation~($0.7\text{--}4.8\%$), and qubit dephasing~($4.2\text{--}10\%$), consistent with the observed Bell-state fidelities.

\section{DISCUSSION}
We have demonstrated deterministic quantum communication between fixed-frequency superconducting transmon qubits using itinerant microwave photons.
By implementing each transfer resonator as two coupled coplanar resonators that simultaneously contribute to the photon emission or absorption process, we achieved resonator bandwidths of approximately 150~MHz for the receiver and 100~MHz for the sender.
Despite the fabrication-related frequency offset of 50~MHz between devices, the broadband design provided a 30-MHz spectral overlap. Across this overlapped frequency range~ ($\omega_{\rm ph}/2\pi = 9.36$--$9.39$~GHz), the photon absorption efficiency remained stable at approximately 95\%.
Both quantum state transfer and remote entanglement generation were successfully demonstrated, achieving process fidelities around 79\% and Bell-state fidelities around 73\%, respectively.
The consistent performance across the frequency range confirms that our approach eliminates the need for flux-tunable circuit elements while maintaining a consistent communication fidelity, establishing a practical pathway toward scalable quantum communication networks based on fixed-frequency qubits. Our approach is particularly advantageous for 3D integrated architectures where magnetic flux lines require specialized implementation~\cite{gargiulo_fast_2021, krasnok_superconducting_2024}.

The distinct resonator designs for the sender and the receiver devices suggest the spectral engineering flexibility enabled by the two-resonator configuration. The design approaches address the fundamental trade-off between operational bandwidth and photon-emission rate: broader bandwidth enables photon emission and absorption across an extended frequency range, while narrower bandwidths, within the limits of the adiabatic condition, make it easier to achieve higher photon-emission rates for faster communication~[Eq.~\eqref{Gamma_f}]. This flexibility allows optimization according to specific application requirements, such as prioritizing broad operational bandwidth or enabling faster communication protocols. Moreover, extending this approach to configurations with more than two coupled resonators could potentially offer enhanced design flexibility for spectral engineering.

The broad operational bandwidth opens new possibilities for frequency-multiplexed quantum communication, where multiple quantum channels operate simultaneously at different carrier frequencies within the same transmission line. A recent theoretical study indicates that microwave waveguides can support multiple concurrent quantum channels when photons are appropriately spaced in frequency, with simulations suggesting the potential for tens of multiplexed channels~\cite{penas_multiplexed_2024}. Combined with our frequency-tunable photon generation method, dense frequency-division multiplexing significantly increases communication capacity. Beyond multiplexed state transfer, the extended frequency-tunable range enabled by the broadband resonator design facilitates frequency-bin encoded quantum information processing~\cite{yang_deterministic_2025} and the generation of frequency-domain cluster states~\cite{osullivan_deterministic_2025, wang_generation_2025} for measurement-based quantum computation. 

The frequency tunability also addresses practical challenges in device operation.
When a strongly coupled TLS defect~\cite{muller_towards_2019} is present near the transfer-resonator frequency, the $\ket{f0}$--$\ket{g1}$ drive can excite the TLS instead of emitting photons, significantly degrading photon emission and absorption. Our approach enables in-situ avoidance of such TLS defects by adaptively selecting photon frequencies that minimize interaction with specific defect resonances, without requiring hardware modifications or device retuning.

Finally, we note that the photon bandwidth in this work ($\kappa_{\rm ph}/2\pi = 2$~MHz) is narrower than those demonstrated in previous studies~\cite{pechal_microwave-controlled_2014, kurpiers_deterministic_2018, storz_loophole-free_2023}. This limitation arises from the adiabatic condition underlying our wave-packet shaping method~\cite{miyamura_generation_2025}. Extending non-adiabatic wave-packet shaping techniques to the off-resonant driving scheme enables faster quantum communication while maintaining the flexibility of frequency-tunable photon generation.

\begin{acknowledgments}
We thank A.~Kulikov, X.~Dai and A.~Hern\'{a}ndez-Ant\'{o}n for the advices for improving the experiments, and R.~Araki, K.~Uchida for the fruitful discussions.
This work was supported in part by the University of Tokyo Forefront Physics and Mathematics Program to Drive Transformation (FoPM), a World-leading Innovative Graduate Study (WINGS) Program, the Ministry of Education, Culture, Sports, Science and Technology (MEXT) Quantum Leap Flagship Program (Q-LEAP) (Grant No.\ JPMXS0118068682), the JSPS Grant-in-Aid for Scientific Research (KAKENHI) (Grant No.\ JP22H04937), and the JST CREST (Grant No.\ JPMJCR23I4).
\end{acknowledgments}

\appendix

\section{EXPERIMENTAL SETUP}\label{app:setup}

\begin{figure*}[t]
    \centering
    \includegraphics{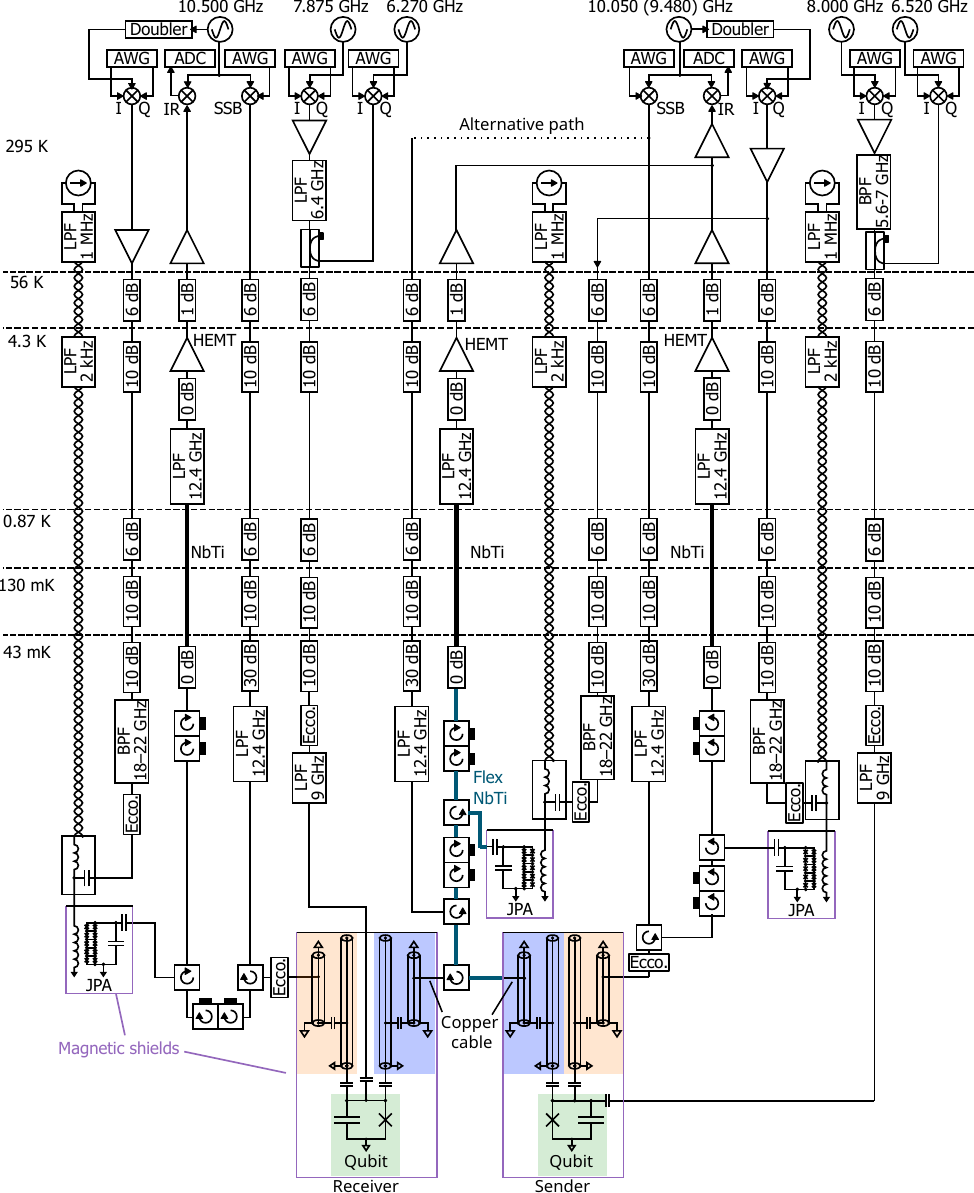}
    \caption{Experimental setup. AWG, arbitrary waveform generator; ADC, analog-to-digital converter; SSB, single-sideband mixer; IR, image-reject mixer; LPF, low-pass filter; BPF, band-pass filter; HEMT, high-electron-mobility transistor; Ecco., eccosorb filter; JPA, Josephson parametric amplifier. The dotted line indicates the alternative signal path used for transfer resonator spectroscopy, which is manually connected by reconfiguring the sender readout line. }
    \label{figure_wiring}
\end{figure*}
The measurement setup used in this work is shown in Fig.~\ref{figure_wiring}.
The sender and receiver devices are connected via a flexible NbTi coaxial cable~(Koaxis, CC086NBTI) with a circulator inserted between them to enable independent characterization of photon emission, propagation loss, and absorption efficiency. The flexible cable uses a niobium-titanium-alloy center conductor with a silver-plated copper-braid shield and PTFE dielectric. 
The NbTi cable has a length of 20~cm. 
A 15-cm copper coaxial cable connects the sender device to the NbTi cable, and another 15-cm copper coaxial cable connects the circulator to the receiver device, resulting in the total cable length of 50~cm between the devices.

To characterize the propagation loss, we performed separate transmission measurements of the cable components at cryogenic temperatures using a microwave switch. The NbTi cable shows negligible attenuation. Each copper cable section with its SMA-SMA connector is estimated to contribute approximately 0.2~dB, giving a combined cable attenuation of approximately 0.4~dB. Including the circulator insertion loss of 0.2~dB (manufacturer specification), the remaining loss of approximately 0.9~dB is attributed to the coaxial-to-printed-circuit-board interfaces at each device, where the coaxial cable is connected to the on-chip coplanar waveguide via solder paste. Both dissipative loss and impedance mismatch at these interfaces may contribute to the observed loss. 
% Because the interfaces cannot be disconnected and reconnected without altering the joint condition, their individual loss contributions cannot be separated from the total measurement. 
% The propagation loss can be reduced by shortening the cable, improving the soldered interfaces, and removing the circulator used here for independent characterization of the emission and absorption processes.

Several components in the setup are shared between different measurement configurations to minimize hardware complexity.
The local oscillator (LO) for the sender qubit readout and for observing the photon waveform is common in both measurement modes.
Consequently, the LO frequency must be alternatively adjusted depending on the measurement configuration: 10.050~GHz for the sender-qubit readout and 9.480~GHz for photon-waveform observation.
Similarly, the AWG, LO, and frequency doubler used for the pump of the JPA are shared between the sender readout chain and the photon observation chain.
The LO frequencies for the $\ket{f0}$--$\ket{g1}$ drives and the qubit control drives are chosen to enable observation of the photon waveform~\cite{ilves_-demand_2020}.
For transfer resonator spectroscopy, the readout line of the sender device is manually reconfigured by connecting it to a dedicated probe line.

The device parameters are listed in Table~\ref{tab_parameters}.
The energy relaxation times listed in the table represent typical values observed during the experiments.
However, we occasionally observe sudden degradation in coherence times.
To provide a complete characterization of the device stability, we report the full range of observed $T_1$ values.
For the sender device, $T_{1,ge}$ ranged from 6 to 31~$\mathrm{\mu s}$, and $T_{1,ef}$ ranged from 2.4 to 11~$\mathrm{\mu s}$.
For the receiver device, $T_{1,ge}$ ranged from 13 to 25~$\mathrm{\mu s}$, and $T_{1,ef}$ ranged from 3.5 to 13.5~$\mathrm{\mu s}$.
These fluctuations are consistent with the reduction in communication fidelity observed at certain photon frequencies, as discussed in Sec.~\ref{sec:comm}.

\begin{table}[t]
    \caption{Device parameters.}
    \centering
    \small  
    \begin{ruledtabular}
    \begin{tabular}{lrr}
    Parameter & Sender & Receiver \\
    \hline\hline
    Qubit frequency $\omega_{eg}/2\pi$ (GHz)& 7.982 & 8.199\\
    Qubit anharmonicity $\alpha/2\pi$ (MHz)& 356 & 352\\
    Readout resonator frequency $\omega_{\mathrm{r}}/2\pi$ (GHz)& 10.14 & 10.50\\
    Readout filter frequency $\omega_{\mathrm{f}}/2\pi$ (GHz)& 10.06 & 10.70\\
    Resonator-resonator coupling $J/2\pi$ (MHz)& 44 &191\\
    Readout filter linewidth $\kappa_{\mathrm{f}}/2\pi$ (MHz)& 81 & 290\\
    Transfer resonator frequency $\omega_{\mathrm{t}}/2\pi$ (GHz)& 9.393 & 9.348\\
    Transfer filter frequency $\omega_{\mathrm{f}}/2\pi$ (GHz)& 9.380 & 9.341\\
    Transfer-resonator coupling $J/2\pi$ (MHz)& 44 & 58\\
    Transfer filter linewidth $\kappa_{\mathrm{f}}/2\pi$ (MHz)& 120 & 137\\
    \makecell[l]{Qubit--readout-resonator\\coupling strength $g_{\mathrm{qr}}/2\pi$ (MHz)}& 151 & 242\\
    \makecell[l]{Qubit--transfer-resonator\\coupling strength $g_{\mathrm{qt}}/2\pi$ (MHz)}& 159 & 142\\
    \hline
    $|g\rangle$--$|e\rangle$ energy-relaxation time $T_{1, ge}$ ($\mu$s)& 20(6) & 18(2)\\
    $|g\rangle$--$|e\rangle$ Ramsey dephasing time $T_{2, ge}^*$ ($\mu$s)& 22(2) & 15(4)\\
    $|g\rangle$--$|e\rangle$ echo dephasing time $T_{2, ge}^{\mathrm{e}}$ ($\mu$s)& 22(6) & 29(5)\\
    $|e\rangle$--$|f\rangle$ energy-relaxation time $T_{1,ef}$ ($\mu$s)& 7.6(2.5) & 8.6(2.8)\\
    $|e\rangle$--$|f\rangle$ Ramsey dephasing time $T_{2,ef}^*$ ($\mu$s)& 7.3(2.3) & 5.9(1.5)\\
    \end{tabular}        
    \end{ruledtabular}
    \label{tab_parameters}
\end{table}

\section{DESIGNING TRANSFER RESONATORS}\label{app:derivation}
\subsection{Derivation of $\Gamma_f$ in case of two resonators}
We derive the photon emission rate $\Gamma_f$ for a transmon qubit coupled to two coplanar resonators connected to a transmission line.
The system Hamiltonian is
\begin{equation}
    \mathcal{H} = \mathcal{H}_{\mathrm{q}} + \mathcal{H}_{\mathrm{r}} + \mathcal{H}_{\mathrm{f}} + \mathcal{H}_{\mathrm{b}} + \mathcal{H}_{\mathrm{qr}}^{\mathrm{c}} + \mathcal{H}_{\mathrm{rf}}^{\mathrm{c}} + \mathcal{H}_{\mathrm{fb}}^{\mathrm{c}},
\end{equation}
where each term represents
\begin{subequations}
\begin{align}
    \mathcal{H}_{\mathrm{q}}/\hbar &= \omega_{eg} b^\dag b + \frac{\alpha}{2}b^\dag b^\dag bb, \\
    \mathcal{H}_{\mathrm{r}}/\hbar &= \omega_{\mathrm{r}}a^\dag a, \\
    \mathcal{H}_{\mathrm{f}}/\hbar &= \omega_{\mathrm{f}} f^\dag f, \\
    \mathcal{H}_{\mathrm{b}}/\hbar &= \int d\omega\,\omega b_{\omega}^\dag b_{\omega}, \\
    \mathcal{H}_{\mathrm{qr}}^{\mathrm{c}}/\hbar &= g (b^\dag a + b a^\dag), \\
    \mathcal{H}_{\mathrm{rf}}^{\mathrm{c}}/\hbar &= J(a^\dag f + a f^\dag), \\
    \mathcal{H}_{\mathrm{fb}}^{\mathrm{c}}/\hbar &= i\sqrt{\frac{\kappa}{2\pi}} \int d\omega\,  (b_{\omega}f^\dag -  b_{\omega}^\dag f).
\end{align}
\end{subequations}
Here, $a$, $f$, $b$ are the annihilation operators for the first resonator, second resonator, and qubit, respectively, and $b_\omega$ represents the annihilation operator for the continuum modes of the transmission line.
The parameters $\omega_{eg}$, $\omega_{\mathrm{r}}$, $\omega_{\mathrm{f}}$ are the qubit transition frequency and resonator frequencies, $\alpha$ is the qubit anharmonicity, $g$ is the qubit-resonator coupling, $J$ is the inter-resonator coupling, and $\kappa$ is the external decay rate of the second resonator.

To analyze the coupled resonator system, we diagonalize $\mathcal{H}_{\mathrm{rf}}=\mathcal{H}_{\mathrm{r}}+\mathcal{H}_{\mathrm{f}}+\mathcal{H}_{\mathrm{rf}}^{\mathrm{c}}$.
This yields new eigenmodes with annihilation operators $a_{\pm}$ and eigenfrequencies
\begin{equation}
    \omega_{\pm} = \frac{\omega_{\mathrm{r}} + \omega_{\mathrm{f}}}{2} \pm \frac{1}{2}\sqrt{\Delta_{\mathrm{fr}}^2 + 4J^2},
\end{equation}
where $\Delta_{\mathrm{fr}}=\omega_{\mathrm{f}}-\omega_{\mathrm{r}}$.
The original operators are related to the eigenmodes through $a = X a_- + Y a_+$ and $f = Z a_- + W a_+$, with transformation coefficients satisfying $X^2 + Y^2 = 1$ and $Z^2 + W^2 = 1$.

This transformation yields effective decay rates for each eigenmode:
\begin{equation}
    \kappa_{-} = \kappa Z^2, \quad \kappa_{+} = \kappa W^2,
\end{equation}
where $\kappa$ is the external decay rate of the outer resonator.

Under microwave driving at frequency $\omega_{\mathrm{d}}$, we work in the rotating frame where the qubit Hamiltonian becomes
\begin{equation}
    \mathcal{H}_{\mathrm{q}}/\hbar = \delta_{eg}b^\dag b + \frac{\alpha}{2}b^\dag b^\dag bb + \frac{\Omega}{2}(b + b^\dag),
\end{equation}
with $\delta_{eg}=\omega_{eg}-\omega_{\mathrm{d}}$ and drive strength $\Omega$.
First-order perturbation theory yields drive-dressed eigenstates ($\ket{\tilde{g}},\,\ket{\tilde{e}},\,\text{and}\,\ket{\tilde{f}}$) and modifies the $|f\rangle$--$|g\rangle$ transition frequency to $\tilde{\delta}_{fg}=2\tilde{\omega}_{eg}+\alpha-2\omega_{\mathrm{d}}$, where $\tilde{\omega}_{eg}$ is the ac-Stark-shifted qubit frequency.

The photon emission rate is obtained by solving the quantum Langevin equations for the eigenmodes $a_{\pm}$:
\begin{subequations}
\begin{align}
    \frac{da_{-}}{dt} &= \left(-i\Delta_{-}-\frac{\kappa_-}{2}\right)a_- -\frac{\sqrt{\kappa_-\kappa_+}}{2}a_+ - \sqrt{\kappa_-}\,b_{\mathrm{in}}, \\
    \frac{da_{+}}{dt} &= \left(-i\Delta_{+}-\frac{\kappa_+}{2}\right)a_+ -\frac{\sqrt{\kappa_-\kappa_+}}{2}a_- - \sqrt{\kappa_+}\,b_{\mathrm{in}},
\end{align}    
\end{subequations}
where $\Delta_{\pm}=\omega_{\pm}-\omega_{\mathrm{d}}$ and $b_{\mathrm{in}}(t)$ is the input field from the transmission line.

Fourier transforming these equations and assuming white noise input yields the frequency-domain response functions:
\begin{equation}
    R_{\pm}(\omega) = \frac{-i\sqrt{\kappa_\pm}\,(\omega-\Delta_{\mp})}{\left[(\omega-\Delta_+)+\frac{i\kappa_+}{2}\right]\left[(\omega-\Delta_-)+\frac{i\kappa_-}{2}\right]+\frac{\kappa_+\kappa_-}{4}}.
\end{equation}

Applying Fermi's golden rule, the total photon emission rate is
\begin{equation}\label{Gamma_f_appendix}
    % \Gamma_f = g_{\text{eff}}^2\left(X^2 |R_{-}(\tilde{\delta}_{fg})|^2+Y^2|R_{+}(\tilde{\delta}_{fg})|^2\right),
    \Gamma_f
  = \frac{g_{\mathrm{eff}}^2 J^2 \kappa}
         {\bigl[(\tilde{\delta}_{fg}-\delta_r)(\tilde{\delta}_{fg}-\delta_f) - J^2\bigr]^2
          + \kappa^2(\tilde{\delta}_{fg}-\delta_r)^2/4},
\end{equation}
where $g_{\text{eff}}=g|\!\bra{\tilde{g}}b\ket{\tilde{f}}\!|$ is the effective coupling strength between the dressed states~\cite{zeytinoglu_microwave-induced_2015}.
We obtain Eq.~\eqref{Gamma_f} in the main text by substituting $\tilde{\delta}_{fg}-\delta_{r(f)}=\omega_{\mathrm{ph}}-\omega_{r(f)}$.

% By explicitly evaluating $|R_{\pm}(\tilde{\delta}_{fg})|^2$ and substituting into Eq.~\eqref{Gamma_f_appendix}, we obtain Eq.~\eqref{Gamma_f} in the main text.

\subsection{Resonator design strategy including photon-emission rate}\label{sec:design}
\begin{figure}[t]
    \centering
    \includegraphics{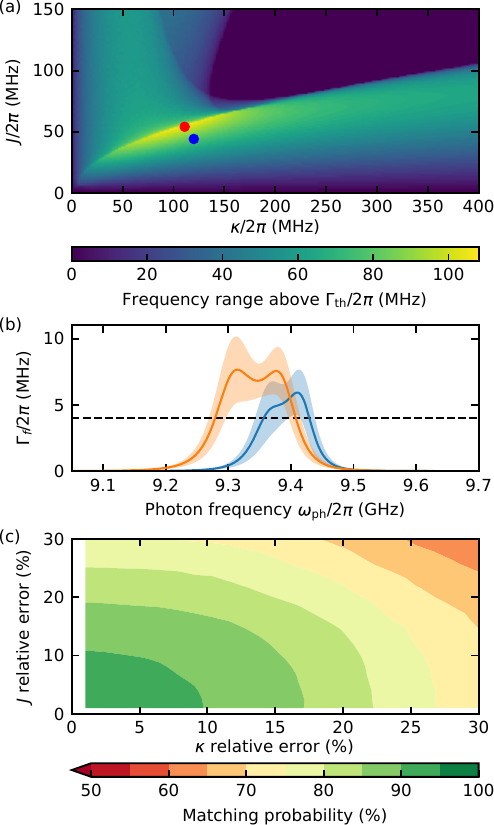}
    \caption{(a)~Frequency range of $\omega_{\mathrm{ph}}$ that satisfies $\Gamma_f(\omega_{\mathrm{ph}})/2\pi>\Gamma_{\mathrm{th}}/2\pi=4$~MHz, evaluated using the measured sender device parameters (Table~\ref{tab_parameters}) and drive strength of $\Omega/2\pi=650$~MHz. The red circle marks the optimal parameters ($\kappa/2\pi = 111$~MHz, $J/2\pi = 55$~MHz) that maximize the operational bandwidth. The blue circle indicates the values of the sender device ($\kappa/2\pi = 120$~MHz, $J/2\pi = 44$~MHz) (b)~Monte Carlo simulation of $\Gamma_f$ as a function of photon frequency. The solid lines show the mean photon-emission rate and the shaded regions show the standard deviation for the sender device (blue) and the receiver device (orange). The horizontal dashed line indicates the threshold $\Gamma_{\mathrm{th}}/2\pi = 4$~MHz. (c) Frequency-matching probability between sender and receiver devices as a function of relative variations in $\kappa$ and $J$.}
    \label{fig:mc}
\end{figure}

Equation~\eqref{Gamma_f} can be used to explore the parameter space of the two-resonator system and identify optimal resonator configurations.
Here, we analyze the dependence of the operational bandwidth on the external decay rate $\kappa$ and the inter-resonator coupling strength $J$ using the measured sender device parameters listed in Table~\ref{tab_parameters} and a drive strength of $\Omega/2\pi=650$~MHz, consistent with the value used in experiments.

We evaluate the frequency range of $\omega_{\mathrm{ph}}$ over which $\Gamma_f(\omega_{\mathrm{ph}})/2\pi\ge\Gamma_{\mathrm{th}}/2\pi=4$~MHz, corresponding to the photon-emission rate used in the experiments ($\kappa_{ph}/2\pi=2$~MHz).
Figure~\ref{fig:mc}(a) shows the frequency range satisfying this criterion as a function of $\kappa$ and $J$.
The optimal parameters form a ridge in the $\kappa$--$J$ plane, reflecting a trade-off inherent to the two-resonator configuration: when $\kappa$ is too small, the resonator bandwidth is narrow, limiting the frequency range over which $\Gamma_f$ remains above the threshold $\Gamma_{\mathrm{th}}$; when $\kappa$ is too large, the spectral density of the inner resonator decreases, reducing $\Gamma_f$ below the threshold.
The parameter set that maximizes the operational bandwidth is $\kappa/2\pi=111$~MHz and $J/2\pi=55$~MHz, which is close to the measured sender device parameters, confirming that the fabricated device operates near the optimal trade-off point. 

\subsection{Monte Carlo simulation of parameter variations}\label{app:MC}
To evaluate the probability of successful frequency matching between independently fabricated sender and receiver devices, we perform Monte Carlo simulations of the transfer-resonator spectrum. We model the complete system using the circuit model corresponding to the fabricated device and calculate the resonator spectrum through electric circuit calculations.

\subsubsection{Estimation of parameter variations}

We assume that the primary source of fabrication uncertainty is the precision limit of photolithography, which we estimate as $\pm 1$~$\mathrm{\mu m}$ for all geometric features. This tolerance affects both coplanar waveguide lengths and lumped-element capacitor geometries. To translate these geometric variations into circuit parameter uncertainties, we perform finite element electromagnetic simulations incorporating the $1\mathrm{\,\mu m}$ tolerance. The capacitance variations obtained from these simulations range from 0.05 to 1.5~fF depending on the capacitor geometry.

To extract the resulting variations in the effective circuit parameters, we calculate the reflection coefficient $S_{11}$ for the complete circuit model including the qubit under different realizations of the geometric tolerances. By fitting these calculated spectra, we obtain estimates of the parameter variations. The resonance frequency of the first resonator shows variations with a standard deviation of approximately 3~MHz, while the second resonator frequency exhibits variations of approximately 20~MHz. The external decay rate and inter-resonator coupling strength both show relative variations of approximately 1\%.
However, the 3-MHz variation in the first resonator frequency appears smaller than values reported in previous studies. To provide a more conservative estimate consistent with reported frequency variations~\cite{valles-sanclemente_post-fabrication_2023, li_optimizing_2023}, we adopt a standard deviation of 10~MHz for this parameter in the subsequent Monte Carlo analysis.

\subsubsection{Monte Carlo simulation}

Using the parameter variations chosen above~(10~MHz for the inner resonator frequency, 20~MHz for the outer resonator frequency, 1\% relative variation for the external decay rate, and 1\% relative variation for the inter-resonator coupling strength), we perform Monte Carlo simulations to evaluate the probability of successful frequency matching between the sender and receiver devices. The nominal parameter values are taken from Table~\ref{tab_parameters}. We generate 2000~parameter sets by adding Gaussian-distributed random variations to both sender and receiver devices independently.

For each parameter set, we calculate the photon-emission rate $\Gamma_f(\omega_{\mathrm{ph}})$ using Eq.~\eqref{Gamma_f}. The drive strength for calculating $\Gamma_f$ is chosen to satisfy the adiabatic conditions $\kappa_+/4 > g_{\mathrm{eff}}$ and $\kappa_-/4 > g_{\mathrm{eff}}$, which ensures that Rabi oscillations are suppressed during the photon emission process. 
% Additionally, the drive strength is kept below $\Omega/2\pi = 750$~MHz, beyond which leakage to higher levels can become significant in the experiments. 
Figure~\ref{fig:mc}(b) shows the mean and standard deviation of $\Gamma_f(\omega_{\mathrm{ph}})$ across the simulated parameter sets for both devices, with the shaded region indicating the mean $\pm$ one standard deviation.

To determine whether a given parameter combination enables successful quantum communication, we require that there exists a common photon frequency $\omega_{\mathrm{ph}}$ of at least 10~MHz over which both devices satisfy $\Gamma_f(\omega_{\mathrm{ph}})/2\pi \geq \Gamma_{\mathrm{th}}/2\pi = 4$~MHz. This 10-MHz requirement reflects the bandwidth of the photon waveform used in the experiments. With these criteria, about 92\% of the simulated device pairs achieve successful frequency matching within a single fabrication batch.

\subsubsection{Sensitivity to large parameter variations}

The estimated 1\% relative variations in the inter-resonator coupling strength $J$ and external decay rate $\kappa$ may be underestimated due to additional sources of uncertainty not fully captured in our geometric tolerance analysis. To assess the design robustness under more pessimistic scenarios, we perform additional Monte Carlo simulations sweeping the relative variations in $J$ and $\kappa$ independently from 1\% to 30\%, while maintaining the frequency variations at 10~MHz and 20~MHz for the first and second resonators, respectively.

Figure~\ref{fig:mc}(c) shows the matching probability as a function of the relative variations in $J$ and $\kappa$. The matching probability remains above 60\% even when both parameters exhibit 30\% relative variations.

\section{Experimental details}\label{app:experiment}
\begin{figure}[t]
    \centering
    \includegraphics{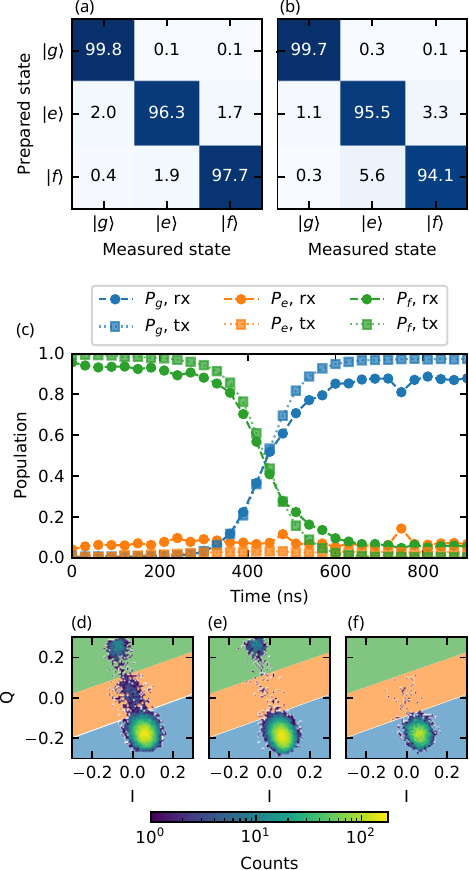}
    \caption{(a)(b)~Readout Assignment matrices for (a)~sender and (b)~receiver devices. (c)~Qubit population after photon emission. Blue, orange, green correspond to the $\ket{g},\,\ket{e}$, and $\ket{f}$ states. Squares are the sender device and the circles are the receiver device. (d)~Single-shot measurement results on the receiver qubit after photon emission. Blue, orange and green areas are assigned to the $\ket{g},\,\ket{e}$, and $\ket{f}$ states. (e)~Single-shot measurement results after photon emission and active reset. (f)~Single-shot measurement results for the qubit prepared in $\ket{g}$ as a reference.}
    \label{fig:qubitpop}
\end{figure}
\subsection{Three-state readout}\label{app:tsr}
We employ dispersive readout to distinguish the three lowest energy levels of the transmon qubit. To improve the readout fidelity, we use JPA in the degenerate mode.

To establish the classification criteria for three-state readout, we first prepare the qubit in each of the three states ($\ket{g}$, $\ket{e}$, $\ket{f}$) and perform single-shot measurements to obtain reference distributions on the IQ plane. We extract the center of each reference distribution by fitting a two-component Gaussian mixture to account for occasional misclassification. Using these three centers as reference points, we partition the IQ plane into three regions by assigning each measurement outcome to the nearest center. 

From the calibration measurements, we construct the assignment probability matrix $R$, where $R_{ij}$ represents the probability of assigning outcome $j$ when the qubit is prepared in state $i$. The assignment matrices for our devices are shown in Figs.~\ref{fig:qubitpop}(a) and~(b). To correct for readout errors, similarly to Refs.~\citenum{kurpiers_deterministic_2018} and~\citenum{magnard_microwave_2020}, we apply matrix inversion to obtain the corrected populations from the measured populations.

\subsection{Qubit population after photon emission}\label{app:qpop}

Figure~\ref{fig:qubitpop}(c) shows the qubit population dynamics during the $\ket{f0}$--$\ket{g1}$ photon-emission pulse for both devices. 
We see that the final $\ket{g}$-state population $P_g$ differs between devices: the sender reaches $P_g=0.95\text{--}0.97$, while the receiver achieves only $P_g=0.85\text{--}0.9$.
Since $P_g$ corresponds to the photon-emission efficiency, this imbalance must be accounted for when estimating photon loss.

To investigate the deviation at the receiver, we perform single-shot measurements following photon-emission process.
We note that these measurements were performed during a separate cooldown from the main quantum communication experiments. 
Figure~\ref{fig:qubitpop}(d) reveals that, while most population resides in $\ket{g}$, residual population persists in the $\ket{e}$ region and in a region corresponding to the $\ket{f}$ and higher states.
Here, our three-state classification assigns all population in the states at or above $\ket{f}$ to a single category, preventing distinction between these levels.

To confirm leakage in $\ket{f}$, we apply active-reset pulses~\cite{magnard_fast_2018} to return the populations in $\ket{e}$ and $\ket{f}$ to $\ket{g}$ before measurement.
Figure~\ref{fig:qubitpop}(e) shows that the population remains in the $\ket{f}$-and-above region after reset, confirming leakage to states above $\ket{f}$.
The leakage population exhibits temporal fluctuations, with values ranging from 0.03 to 0.1.
The origin of this leakage remains unclear, though possible mechanisms include higher-order transitions or coupling to spurious modes during the emission process~\cite{dai_spectroscopy_2025}.

\subsection{Photon absorption dynamics}\label{app:qdynamics}
\begin{figure}[t]
    \centering
    \includegraphics{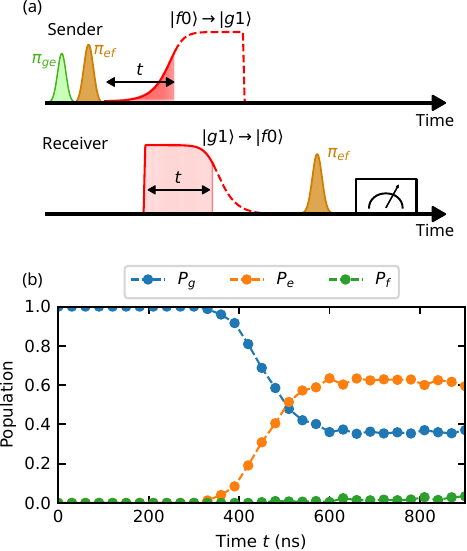}
    \caption{(a)~Pulse sequence for measuring the photon absorption dynamics at the receiver. (b)~Receiver qubit population as a function of drive duration at $\omega_{\mathrm{ph}}/2\pi=9.39$~GHz. Blue, orange, green correspond to the $\ket{g},\,\ket{e}$, and $\ket{f}$ states, respectively.}
    \label{fig:qubitdynamics}
\end{figure}
\begin{table}[t]
    \caption{Receiver qubit population after absorption.}
    \centering
    \small  
    \begin{ruledtabular}
    \begin{tabular}{lccc}
    $\omega_{\mathrm{ph}}/2\pi$~(GHz) & $P_g$ & $P_e$ & $P_f$ \\
    \hline
    9.36 & 0.355 & 0.595 & 0.051\\
    9.37 & 0.396 & 0.593 & 0.011\\
    9.38 & 0.350 & 0.647 & 0.003\\
    9.39 & 0.371 & 0.595 & 0.034\\
    \end{tabular}        
    \end{ruledtabular}
    \label{tab_dynamics}
\end{table}
To characterize the photon absorption process at the receiver, we measure the receiver qubit population during the absorption protocol.
The pulse sequence is shown in Fig.~\ref{fig:qubitdynamics}(a).
The sender qubit is prepared in $\ket{e}$, transferred to $\ket{f}$ via a $\pi_{ef}$ pulse, and a single photon is emitted at frequency $\omega_{\mathrm{ph}}$ following the $\ket{f0}$--$\ket{g1}$ drive. 
At the receiver, the time-reversed drive absorbs the photon, and a subsequent $\pi_{ef}$ pulse maps the $\ket{f}$ population to $\ket{e}$. The delay between the emission and absorption drives is chosen to maximize the receiver $\ket{f}$-state population as explained in the main text.
Figure~\ref{fig:qubitdynamics}(b) shows the receiver qubit population as a function of the absorption drive duration at $\omega_{\mathrm{ph}}/2\pi=9.39$~GHz. The population in $\ket{e}$ increases monotonically and saturates, reflecting the progressive capture of the incoming photon.
Table~\ref{tab_dynamics} summarizes the final receiver qubit population after the full absorption protocol at each photon frequency. 
The measured $\ket{e}$-state populations are consistent with expectations from the independently measured photon loss ($L\approx 0.29$) and absorption efficiency ($\eta \approx 0.95$), accounting for qubit relaxation during the protocol and gate imperfections.

\subsection{State tomography}
Quantum state tomography is performed using the single-shot readout method described in Appendix~\ref{app:tsr}. We apply tomography gates prior to measurement and reconstruct density matrices via maximum-likelihood estimation~\cite{kurpiers_deterministic_2018}.

For process tomography, qutrit-level measurements on both the sender and receiver are necessary to characterize the population remaining in $|f\rangle$ after state transfer. 
% The process matrix is obtained by projecting the reconstructed qutrit density matrices onto the qubit subspace and applying linear inversion.
For each output state, we first reconstruct the full qutrit density matrix using maximum likelihood estimation~(MLE), which guarantees positivity and unit trace.
The qubit-subspace density matrix is then obtained by extracting the upper-left $2\times2$ block without renormalizing the trace, so that any residual population in $\ket{f}$ is reflected as a trace deficit.
The process matrix $\chi$ is obtained from these qubit-subspace density matrices using a semi-definite program that enforces $\chi\ge0$.
% For Bell-state analysis, we perform two-qutrit tomography and extract the qubit subspace to obtain $\rho$ from the full two-qutrit density martix~$\rho_{3\otimes 3}$~\cite{kurpiers_deterministic_2018,magnard_microwave_2020}.
For Bell-state analysis, we similarly reconstruct the full two-qutrit density matrix via MLE and extract the two-qubit subspace corresponding to $\ket{gg},\,\ket{ge},\,\ket{eg},$ and $\ket{ee}$ without renormalizing the trace.

\allowdisplaybreaks
\subsection{Numerical simulation of quantum communication}\label{app:sim}
Numerical simulations are performed using the cascaded-system formalism~\cite{kurpiers_deterministic_2018}.
We approximate the two-pole transfer resonators as single-pole resonators for computational simplicity.
The effective Hamiltonian and master equation for the cascaded system are
\begin{align}
    \mathcal{H}_{\mathrm{eff}} &= -\frac{\alpha_{\mathrm{tx}}}{2}b_{\mathrm{tx}}^\dag b_{\mathrm{tx}} + \frac{\alpha_{\mathrm{tx}}}{2} b_{\mathrm{tx}}^\dag b_{\mathrm{tx}}^\dag b_{\mathrm{tx}} b_{\mathrm{tx}} \nonumber\\
    &\quad -\frac{\alpha_{\mathrm{rx}}}{2}b_{\mathrm{rx}}^\dag b_{\mathrm{rx}} + \frac{\alpha_{\mathrm{rx}}}{2} b_{\mathrm{rx}}^\dag b_{\mathrm{rx}}^\dag b_{\mathrm{rx}} b_{\mathrm{rx}} \nonumber\\
    &\quad + g_{\mathrm{eff}}^{\mathrm{tx}}(t)(a_{\mathrm{tx}} b_{\mathrm{tx}}^\dag b_{\mathrm{tx}}^\dag + \mathrm{h.c.})/\sqrt{2} \nonumber\\
    &\quad + g_{\mathrm{eff}}^{\mathrm{rx}}(t)(a_{\mathrm{rx}} b_{\mathrm{rx}}^\dag b_{\mathrm{rx}}^\dag + \mathrm{h.c.})/\sqrt{2} \nonumber\\
    &\quad - i\frac{\sqrt{\eta_{\mathrm{sim}}\kappa_{\mathrm{tx}}\kappa_{\mathrm{rx}}}}{2}(a_{\mathrm{tx}}^\dag a_{\mathrm{rx}} - \mathrm{h.c.})
\end{align}
% \newpage
\begin{equation}
\begin{aligned}
    &\dot{\rho} = -i[\mathcal{H}_{\mathrm{eff}},\, \rho] \\
    &+ \mathcal{D}[\sqrt{\eta_{\mathrm{sim}}\kappa_{\mathrm{tx}}}\,a_{\mathrm{tx}} + \sqrt{\kappa_{\mathrm{rx}}}\,a_{\mathrm{rx}}]\rho + \mathcal{D}[\sqrt{(1-\eta_{\mathrm{sim}})\kappa_{\mathrm{tx}}}\,a_{\mathrm{tx}}]\rho \\
    &+\sum_{\mu=\mathrm{tx, rx}} \left(
    \gamma_{1, ge}^\mu\mathcal{D}[|g\rangle\!\langle e|_\mu]\rho
    +\gamma_{1, ef}^\mu \mathcal{D}[|e\rangle\!\langle f|_\mu]\rho \right.\\
    &\left.+\gamma_{\phi, ge}^\mu \mathcal{D}[|e\rangle\!\langle e|_\mu-|g\rangle\!\langle g|_\mu]\rho 
    +\gamma_{\phi, ef}^\mu \mathcal{D}[|f\rangle\!\langle f|_\mu-|e\rangle\!\langle e|_\mu]\rho \right),
\end{aligned}
\end{equation}
where $\mathcal{D}[L]\bullet=L\bullet L^\dag - \{L^\dag L,\,\bullet\}/2$ is the dissipation superoperator, $\mu\in\{\mathrm{tx,\,rx}\}$ labels the device, $\kappa_{\mathrm{tx(rx)}}$ is the transfer-resonator linewidth, $\eta_{\mathrm{sim}}$ is the effective photon-loss parameter used in the simulation, $\gamma_{1,ge(ef)}^\mu = 1/T_{1,ge(ef)}^\mu$ is the energy-relaxation rate, and $\gamma_{\phi,ge(ef)}^\mu = 1/T_{2,ge(ef)}^{\mu} - 1/2T_{1,ge(ef)}^\mu$ is the pure dephasing rate.
For the simulations, we use $\kappa_{\mathrm{tx}}/2\pi=150$~MHz and $\kappa_{\mathrm{rx}}/2\pi=200$~MHz.

To quantify individual error contributions, we perform simulations under controlled conditions. We isolate photon-loss and absorption-inefficiency contributions by comparing error-free simulations with simulations including only these loss mechanisms (modeled through the loss parameter $\eta$). The absorption inefficiency arises from waveform mismatch between emission and absorption processes; in our simulations, we assume perfect waveform matching and incorporate the measured absorption inefficiency directly into $\eta_{\mathrm{sim}}$ such that $\eta_{\mathrm{sim}}=\eta(1-L)$. Under these assumptions, the specific values of $\kappa_{\mathrm{tx}}$ and $\kappa_{\mathrm{rx}}$ have negligible impact on the simulation results. Decoherence contributions from $T_1$ and $T_\phi$ are estimated by comparing simulations with photon loss (but with infinite coherence times) against simulations with finite $T_1$ and/or $T_\phi$.
\\ \\ \\ \\ \\
% \clearpage
\FloatBarrier
\nocite{*}
\bibliography{PhotonCommunication}  % must use BibTeX, not BibLaTeX

\end{document}